\RequirePackage{fix-cm}
\documentclass{svjour3}                     % onecolumn (standard format)
\smartqed  % flush right qed marks, e.g. at end of proof
\usepackage{graphicx}
\usepackage{booktabs}
\usepackage{multirow}
\usepackage{float}
\usepackage{subcaption}
\usepackage{rotating}
\usepackage{booktabs,caption}
\usepackage[flushleft]{threeparttable}
\begin{document}

\title{Gamma attenuation parameters of some borate glasses%\thanks{Grants or other notes
}


\author{M.S. Al-Buriahi %\and Amani Alalawi   %etc.
}


\institute{M.S. Al-Buriahi  \at
          Department of Physics, Sakarya University, Sakarya, Turkey \\
              \email{mohammed.al-buriahi@ogr.sakarya.edu.tr }           %  \\
%\and
 %         Amani Alalawi \at
  %            Department of Physics, Umm AL-Qura University, Makkah, Saudi Arabia
}

\date{Received: date / Accepted: date}

\maketitle

\begin{abstract}
In this study we have reported the borate-based glasses in the form of $(75-x)B_2O_3-xBi_2O_3-10Na_2O-10CaO-5Al_2O_3$ (0 $\leq$ x $\leq$ 25 mol$\%$) for gamma-ray and neutron shielding applications, and for this purpose mass attenuation coefficients ($\mu_m$) of these glasses have been investigated using Geant4 code and XCOM program in the energy range between 1 keV and 10 MeV. A good agreement between the simulation results and computed values was observed. The obtained $\mu_m$ values were then used to calculate the other related shielding parameters such as half value layer ($HVL$), mean free path (MFP) and effective atomic numbers ($Z_{eff}$). Exposure buildup factors of the glasses were also evaluated by using the geometrical progression (G-P) method. Additionally, neutron effective removal cross-sections of each selected glass have been computed. The results indicate that the $Bi_2O_3$ content in the glass sample affects the radiation shielding properties. It can be concluded that the glass coded as F offers superior gamma-ray shielding properties, while the glass coded as D offers better neutron shielding features among the other samples.

\keywords{Gamma attenuation \and Geant4 \and XCOM \and borate glass}
\end{abstract}

\section{Introduction}
Radiation shielding by using heavy metal oxide (HMO) glasses is a significative field of study~\cite{al2019radiation,al2019study}. Its interest has mushroomed dramatically in the wake of nuclear accidents such as Chernobyl disaster which caused deleterious effects for human being. The conventional materials such as lead-based materials and concretes are inexpensive, high durability and can be cast to any desired shape easily. Nevertheless, there are still limitations associated with their use because they have several disadvantages like being nontransparent shielding materials, cracks formation and water permeability~\cite{kurudirek2018effect,al2020comparison}. Further, the toxicity and hazardous nature of lead (Pb) to environment and human health are serious concerns to replace lead by other high atomic number element such as Te, Ba, W, Bi, and Gd etc. This triggered much attention to synthesize numerous materials (e.g. concretes, alloys, polymers, glasses...ect.) with different structures in order to achieve environmentally friendly radiation shielding candidates ~\cite{bel2019radiation,alfaify2018facile,mabrouk2014comparative,rani2018role,wagh2018gamma,mahmoud2018fabrication,ersundu2018heavy}. Consequently,  efforts should be made to investigate the shielding properties of these manufactured materials to use in real applications. Among those materials the glasses can be the preferable ones due to their features of being transparent for the light and their capacity to be a host for almost all the elements. For example, the addition heavy metal oxide (HMO) in the structure of the glass leads to rise the $Z_{eff}$ and $N_{eff}$ of the samples. This makes them appropriate to use as a superior shielding for ionizing radiations. 

Literature has been enriched by several authors who provided successful studies on the radiation shielding properties for several glass systems~\cite {bagheri2018determination,al2020investigation,al2020mechanical}. The existence of HMO in these glasses is an indispensable requirement to obtain interesting properties for various applications. In this perspective, the glasses doped with bismuth oxide (as one of HMO family) have fascinated attention of many authors because of their structural and optical properties that have importance in optoelectronic and shielding applications. Early, the mass attenuation coefficients for the glass system $xBi_2 O_3$ $(1-x)B_2 O_3$ (0.30 $\leq$ x $\leq$ 0.55 mol$\%$) were experimentally determined and found to be constant with bismuth oxide $(Bi_2O_3)$ concentration at the energies of 356, 662, 1173 and 1332 keV \cite{singh2002gamma}. The effective atomic numbers for the glasses in the forms: $CaO-SrO-B_2 O_3$, $PbO-B_2 O_3$, $Bi_2O_3-B_2O_3$ , and $PbO-Bi_2O_3-B_2O_3$ were calculated at energies between 1 keV and 1 GeV, the results concluded that the bismuth oxide is better than lead oxide in order to increase the shielding capacity\cite{manohara2009photon}. Recently, MCNPX code and XCOM program were utilized to investigate the nuclear shielding properties for different borate and tellurite glass systems\cite{al2020mechanicalA}. The shielding properties of $TeO2-ZnO-NiO$ glass system  were evaluated by Geant4 and the results were compared with MCNPX and XCom program. The lithium zinc bismuth borate glass systems were studied against gamma ray and fast neutrons by using XCOM program and MCNP5 simulations\cite{al2020polarizability}. These glasses showed superior shielding features comparing to commercial glasses and ordinary concrete. The borosilicate glass in the composition $50$ $Bi_2O_3–15$ $B_2O_3–(35-x)$ $ZnO–(x)$ $Li_2O$ ($x=0, 5, 10, 15, 20$  mol $\%$) were prepared and the authors focused on the effect of bismuth oxide to improve the shielding efficiency comparing to different glasses and concretes\cite{kurudirek2018effect}. Very recently, the borate-based glasses use as attractive materials for optoelectronics, nonlinear optical, and photonic instruments like; switches, waveguide amplifiers, and electrical storage devices~\cite{shinozaki2019synthesis,ciceo2007ft}. This has encouraged us to test the shielding properties of these glasses for potential protection applications against ionizing radiation. 

In the present work, the gamma and neutron radiation shielding properties of some borate-based glasses have been investigated over a wide range of energies from 1 keV to 10 MeV. The values of $\mu_m$ have been determined by using Geant4 simulations and XCOM program. The $\mu_m$ values were then used to calculate $Z_{eff}$, HVL, and MFP of the present glasses. The exposure buildup factors have been calculated for the photon energies up to 15 MeV and penetration depths up to 40 mfp. Moreover, fast neutron removal cross sections have been calculated for each selected glass. The shielding properties of the selected glasses were compared in terms of MFP with conventional concretes and other HMO glasses. This study on shielding properties of the borate-based glasses would be useful in various radiation protection applications such as exposure control.

 \section{Materials and methods}
 
 The $(75-x)B_2O_3-xBi_2O_3-10Na_2O-10CaO-5Al_2O_3$ (0 $\leq$ x $\leq$ 25 mol$\%$) were investigated by means of their radiation shielding competence. These glasses were coded as A, B, C, D, E, and F according to increase of $Bi_2O_3$ content from 0 to 40 mol$\%$, respectively. The weight fractions of the elements in GSC-X glass system along with the glass density are summarized in Table~\ref{tab1}. 
 
The gamma attenuation parameters of the glasses were investigated by using Geant4 toolkit for several energies~\cite{agostinelli2003geant4}. The detailed information about the simulation setup and procedures can be found in Refs.~\cite{al2019mass,al2020mass,al2020gamma}. The outcomes of the simulations were compared with the theoretical predictions of XCOM program~\cite{berger1999xcom}. Moreover, we calculated the effective atomic number for each glass system as described by Hine~\cite{hine1952effective}. For this calculation part, one can find all the reqiured equations in Refs.~\cite{manohara2008effective,al2019investigation}.

\section{Results and discussion} 
Chemical compositions and the densities of the studied borate-based glasses are listed in Table~\ref{tab1}. It is clear that the densities of these glasses vary from a lower value (2.250 $g$ $cm^{-3}$ for the glass A) to a higher value (5.035 $g$ $cm^{-3}$ for the glass F), due to the increment of $Bi_2O_3$ from 0 to 25 mol$\%$.

\begin{figure}[H]
\begin{center}
\includegraphics[width=0.8\textwidth,natwidth=610,natheight=642]{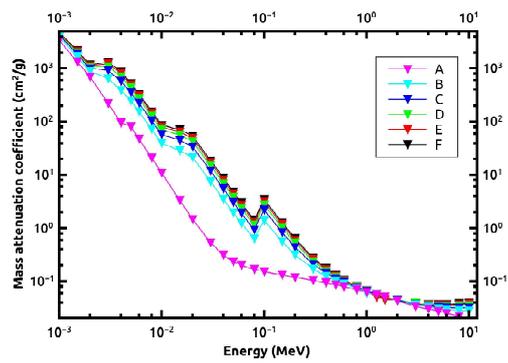}
\caption{Mass attenuation coefficient ($\mu_m$) for borate-based glasses by using Geant4 simulations in the photon energy region from 0.001 MeV to 10 MeV.}
\label{fig1}
\end{center}
\end{figure}

 The simulated results of $\mu_m$ for the studied glass samples are shown in Fig.~\ref{fig1}. It is clearly seen that the values of $\mu_m$ rely on the photon energy and the constituent elements of the glass. Such that the mass attenuation coefficent decrease with the increase of photon energy and increase with the increase of $Bi_2O_3$. The energy dependence of $\mu_m$ is reflected the partial photon interactions. For example, the swift decrease of $\mu_m$ values in the low energy region (0-400 keV) is attributted to the photoelectric interaction that relative to $E^3$. At the intermediate energy region (400 keV-3 MeV), one can notice a slight decrease of $\mu_m$ due to the dominance of Compton scattering which has a poor cross section according to its independent on atomic number of the sample (say the glass). Therefore, we found that all studied glasses have the same value of $\mu_m$ at 1 MeV and 2 MeV. In the high energy region (3-10 MeV), the pair production is found to be predominant process. It is worth mentioning that the pair production occur when the photon energy is higher than 1.022 MeV, but it is an important source of attenuation at energies around 10 MeV. Therefore, we observed an increase in $\mu_m$ values at 10 MeV for glass F with 25 $\%$ mol of $Bi_2O_3$ (e.g. Bi is high-Z element and the cross section of the pair production process is directly related to $Z^2$). 
 
\begin{figure}[H]
\begin{center}
\includegraphics[width=0.8\textwidth,natwidth=610,natheight=642]{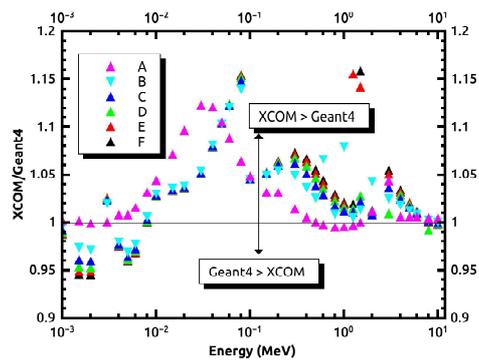}
\caption{Comparison between the values of $\mu_m$ obtained by Geant4 and XCOMfor borate-based  glasses.}
\label{fig2}
\end{center}
\end{figure}

Fig.~\ref{fig2} shows the comparison of $\mu_m$ obtained by Geant4 package and XCOMprogram. From Fig.~\ref{fig2}, it is observed that there is a good agreement between the simulated results and calculated values while the maximum discrepancies up to 15 $\%$ were observed at 80 keV and 2 MeV. At low energies, the results of Geant4 simulation come higher than those of XCOM database due to the essential inconsistency between the cross section data in XCOM and the cross sections of EPDL97\cite{cullen1997epdl97}, on which the Geant4 low energy parameterised model is based. According to EPDL97, it is not recommended to use this data for calculation the photon transport at low energies. At intermedite and high energies the values of XCOM are higher than those of Geant4 due to the molecular bonding and solid-state effects have been neglected in the cross section calculations \cite{berger1987xcom}. For example, the difference between the results of XCOM and Geant4 increase as the $Bi_2O_3$ increases in the glass sample. Therefore, in this case we recommend to use the Geant4 results. Usually, it is believed that the uncertainty in the values of XCOM is 5 $\%$ \cite{cullen1989tables}, and the uncertainty of Geant4 simulation  is 3 $\%$ \cite{amako2005comparison}. It seems that for unpretentious physic models as transmission problem as the present study, the chosen of parameter by any Monte Carlo platform affects the final results of the simulation. The significant differences (up to 14.431 $\%$) between Geant4 and MCNPX have been observed at low energies (0.662 MeV ), while the differences become smaller at high energies. These divergences may be occured due to the different physical models were used to underlie the photon cross-sections, however the multiple scattering for the photon at high energies reduce the discrepancies of the attenuation parameters that obtained by either Geant4 or MCNPX. 

\begin{figure}[H]
\begin{center}
\includegraphics[width=0.8\textwidth,natwidth=610,natheight=642]{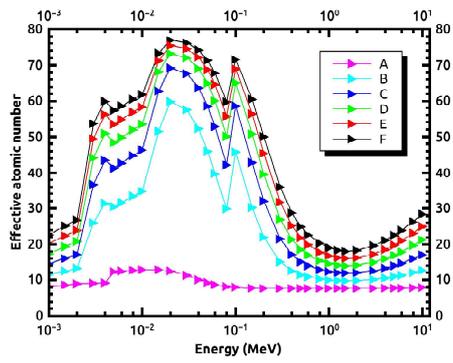}
\caption{Effective atomic number ($Z_{eff}$) for borate-based  glasses in the energy
range 0.001 MeV to 10 MeV.}
\label{fig3}
\end{center}
\end{figure}

\begin{figure}[H]
\begin{center}
\includegraphics[width=0.8\textwidth,natwidth=610,natheight=642]{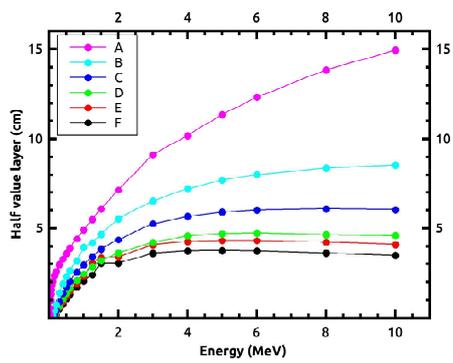}
\caption{Half value layer (HVL) as a function of photon energy for borate-based  glasses.}
\label{fig4}
\end{center}
\end{figure}

\begin{figure}[H]
\centering
\begin{subfigure}{.6\textwidth}
  \centering
  \includegraphics[width=1.1\linewidth]{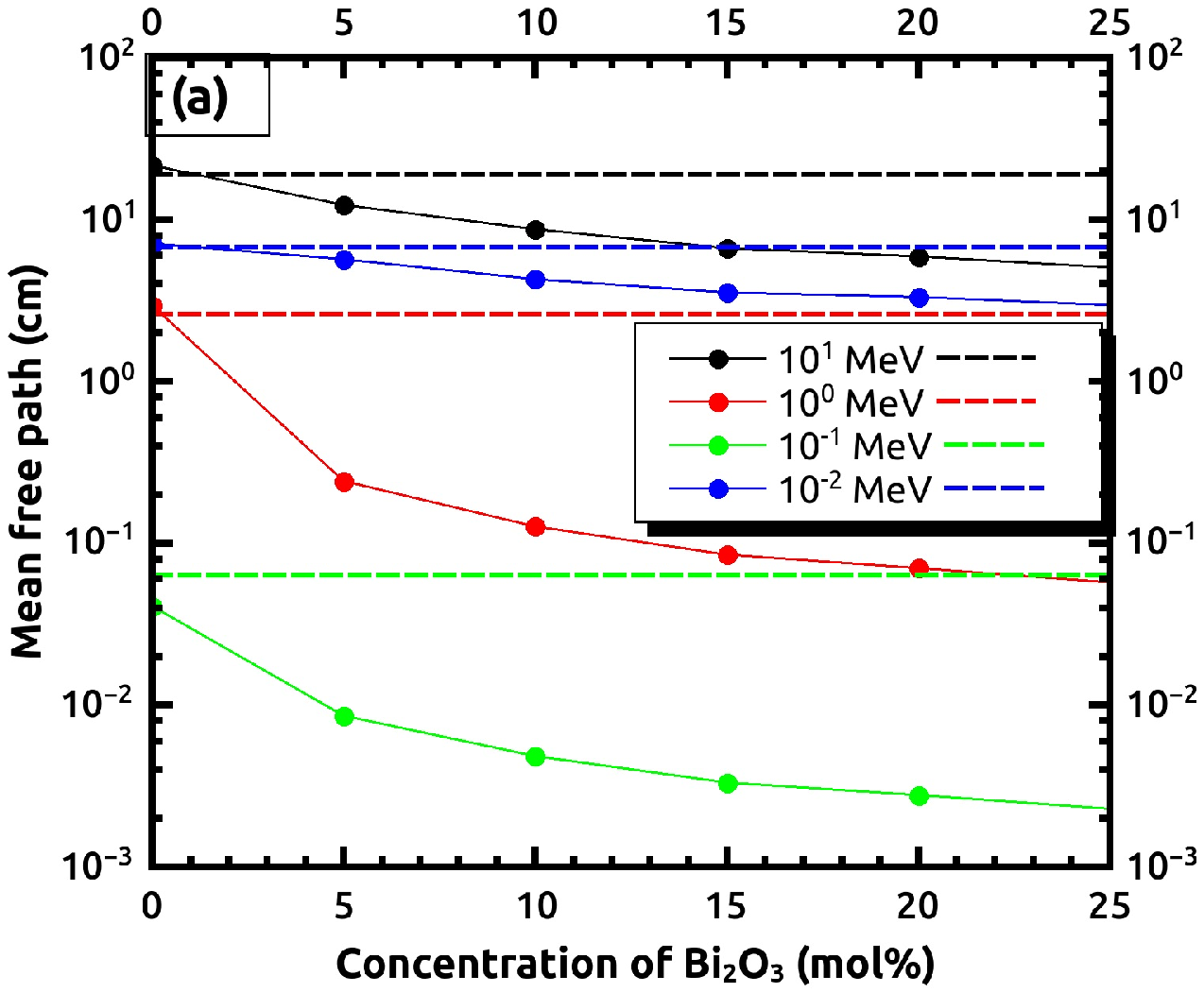}
  \label{fig:sub1}
\end{subfigure}%
\begin{subfigure}{.6\textwidth}
  \centering
  \includegraphics[width=1.08\linewidth]{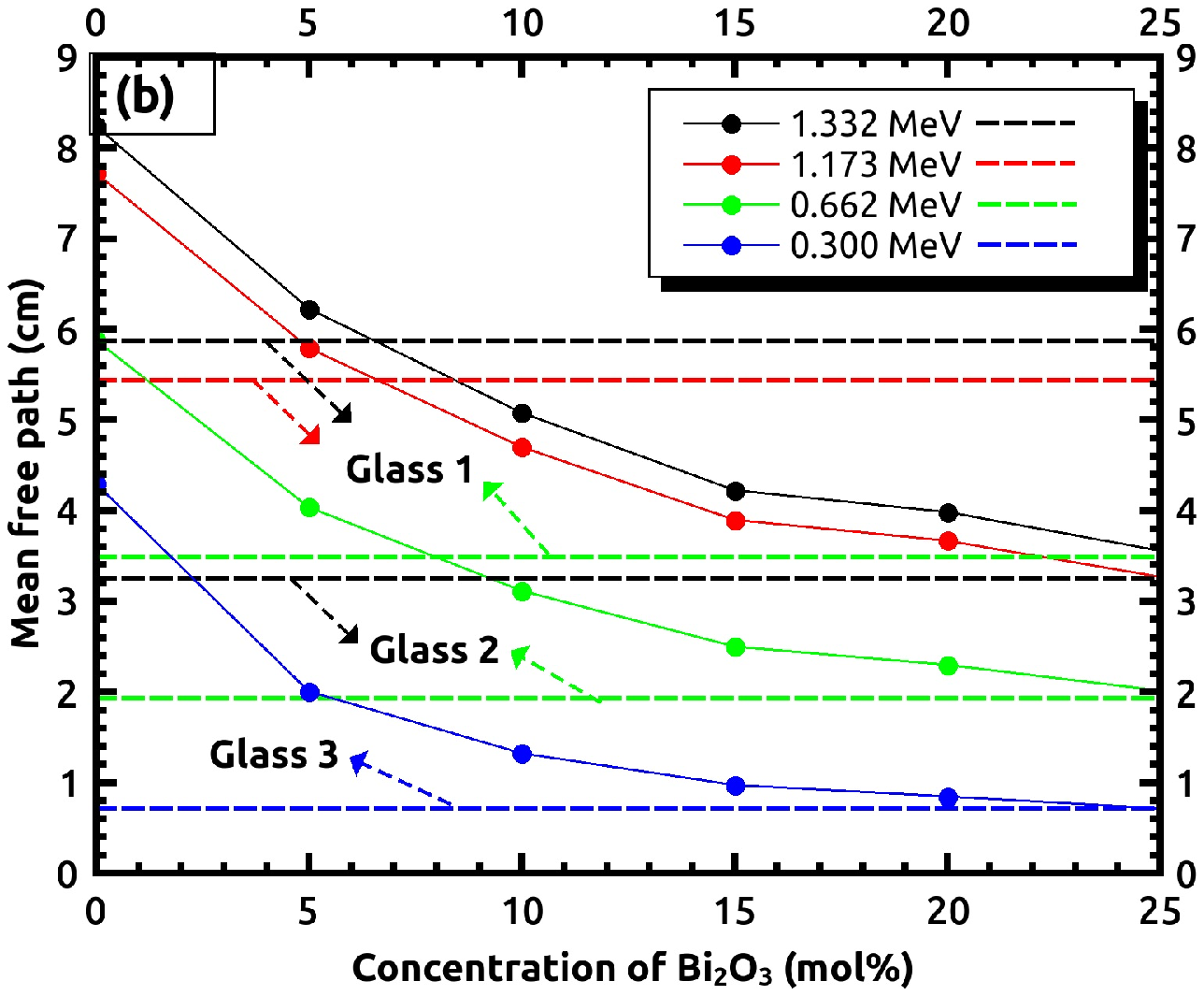}
  \label{fig:sub2}
\end{subfigure}
\caption{Mean free path (MFP) for borate-based  glasses with different concentration of $Bi_2O_3$, (a) for comparison the glass samples with ordinary concrete\cite{bashter1997calculation}, and (b) for comparison the glass samples with glass 1\cite{kurudirek2018effect} and glass 2\cite{singh2002gamma} and glass 3.}
\label{fig5}
\end{figure}

\begin{figure}[H]
\centering
\begin{subfigure}{.6\textwidth}
  \centering
  \includegraphics[width=1.08\linewidth]{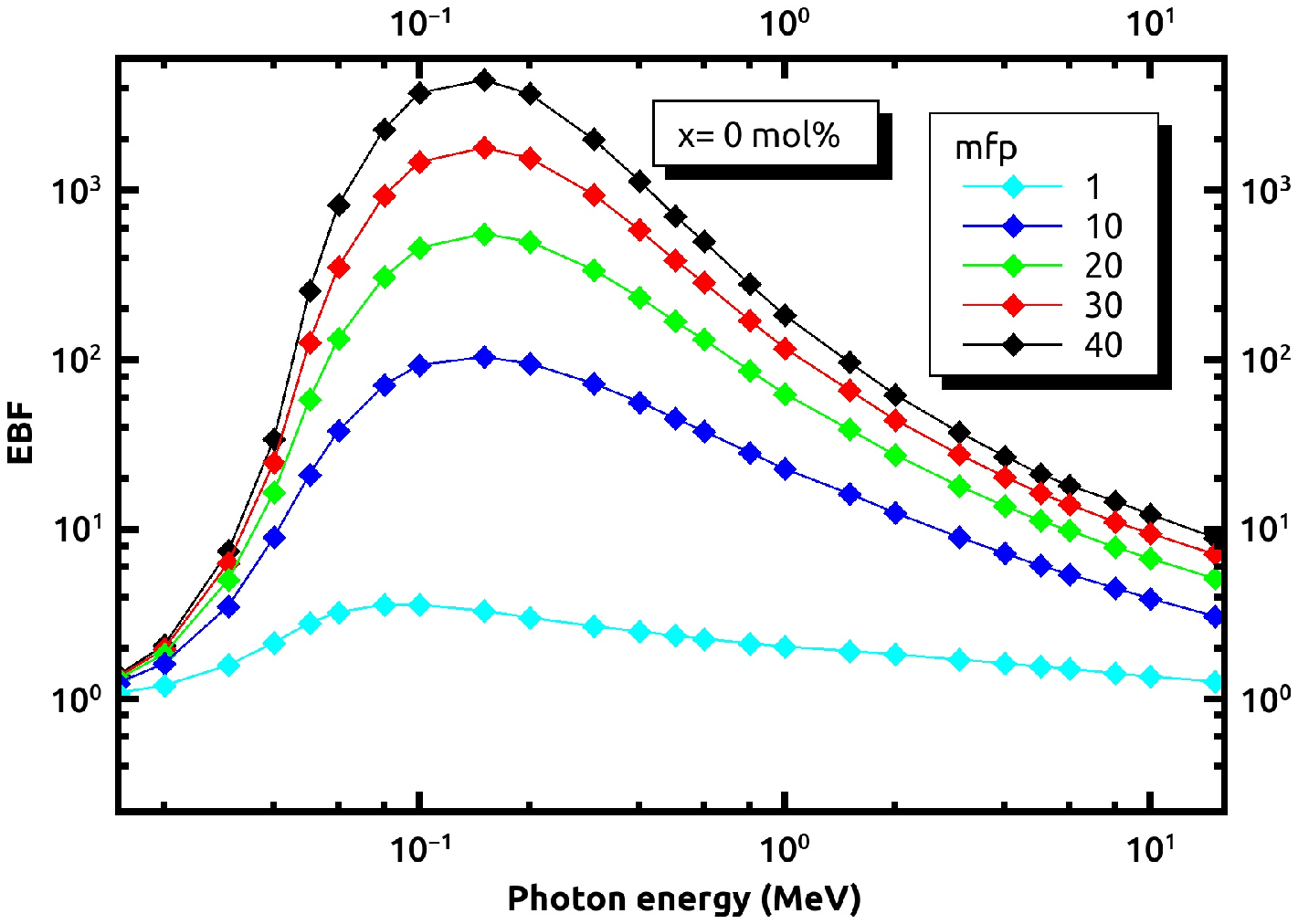}

  \label{fig:sub3}
\end{subfigure}%
\begin{subfigure}{.6\textwidth}
  \centering
  \includegraphics[width=1.08\linewidth]{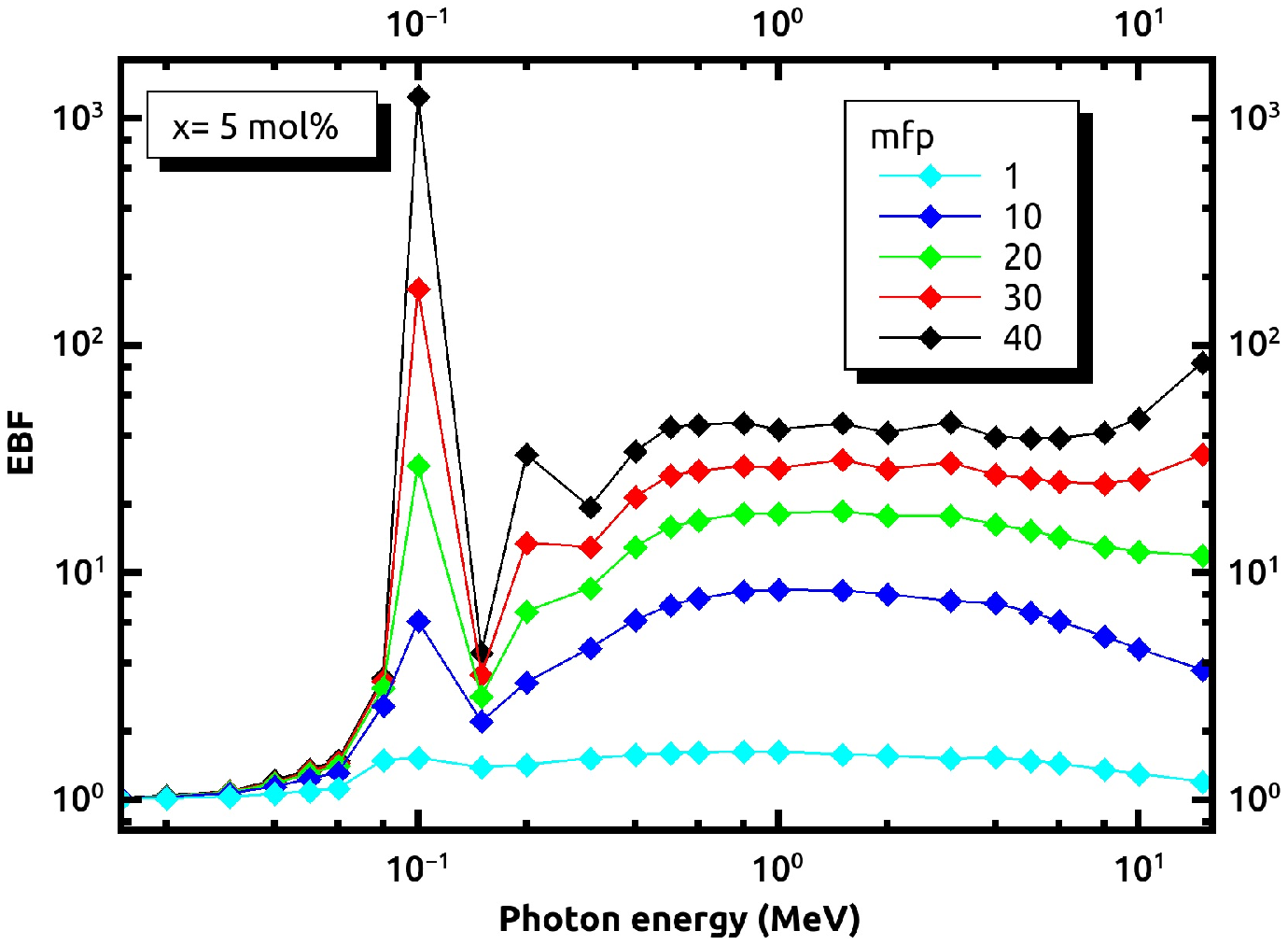}
  
  \label{fig:sub4}
\end{subfigure}
\begin{subfigure}{.6\textwidth}
  \centering
  \includegraphics[width=1.08\linewidth]{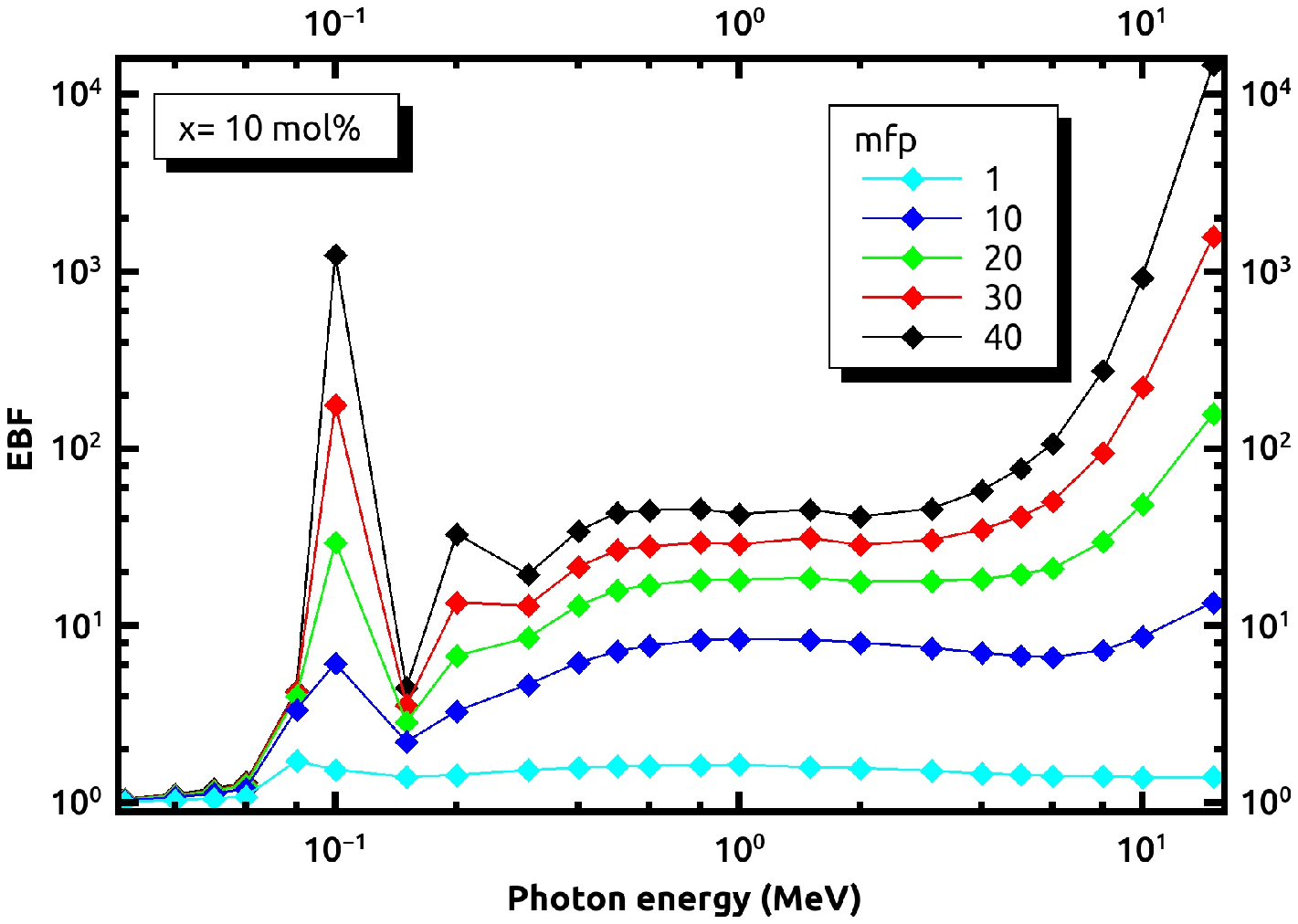}
  
  \label{fig:sub5}
\end{subfigure}%
\begin{subfigure}{.6\textwidth}
  \centering
  \includegraphics[width=1.08\linewidth]{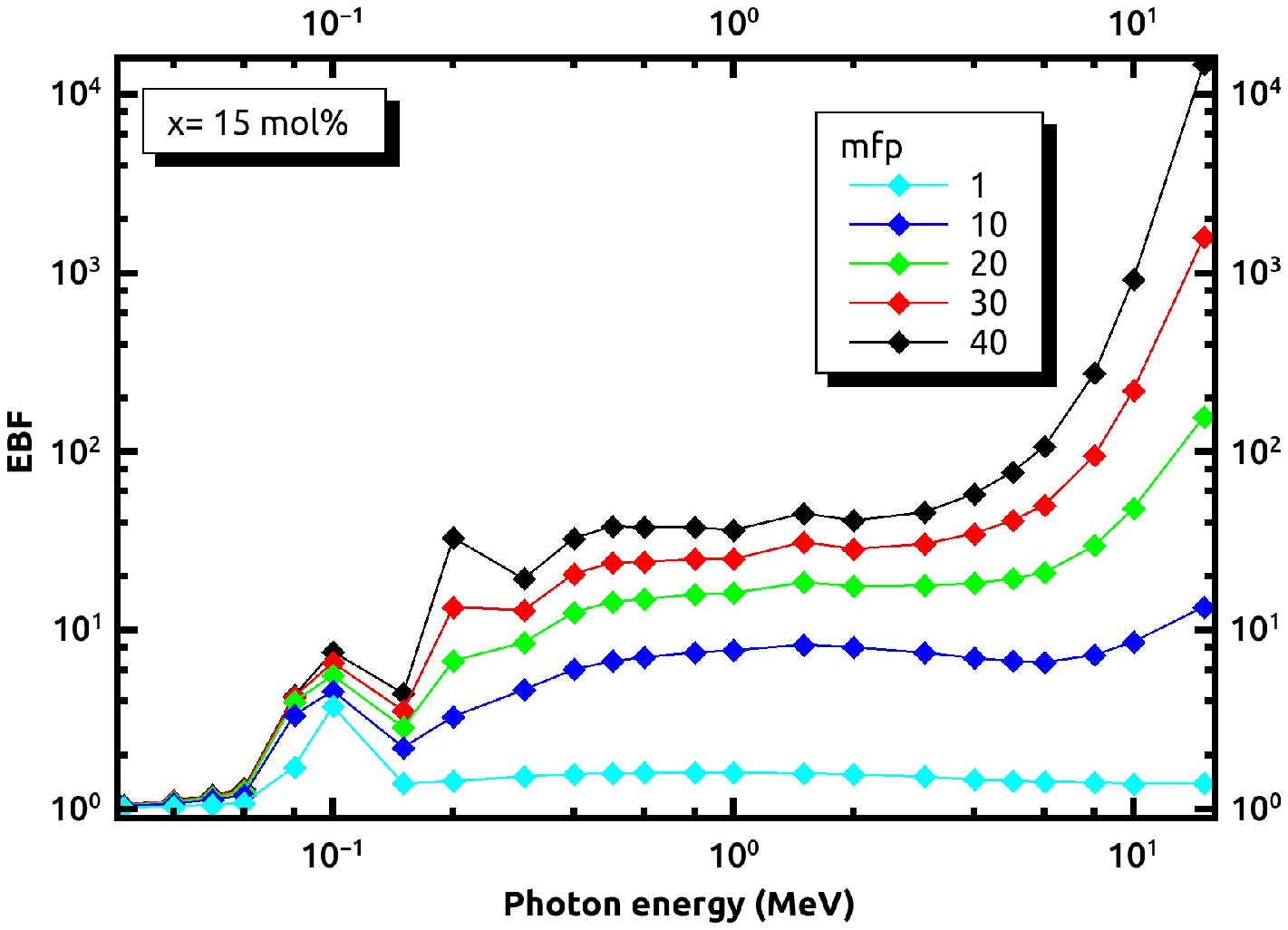}
  
  \label{fig:sub6}
\end{subfigure}
\begin{subfigure}{.6\textwidth}
  \centering
  \includegraphics[width=1.08\linewidth]{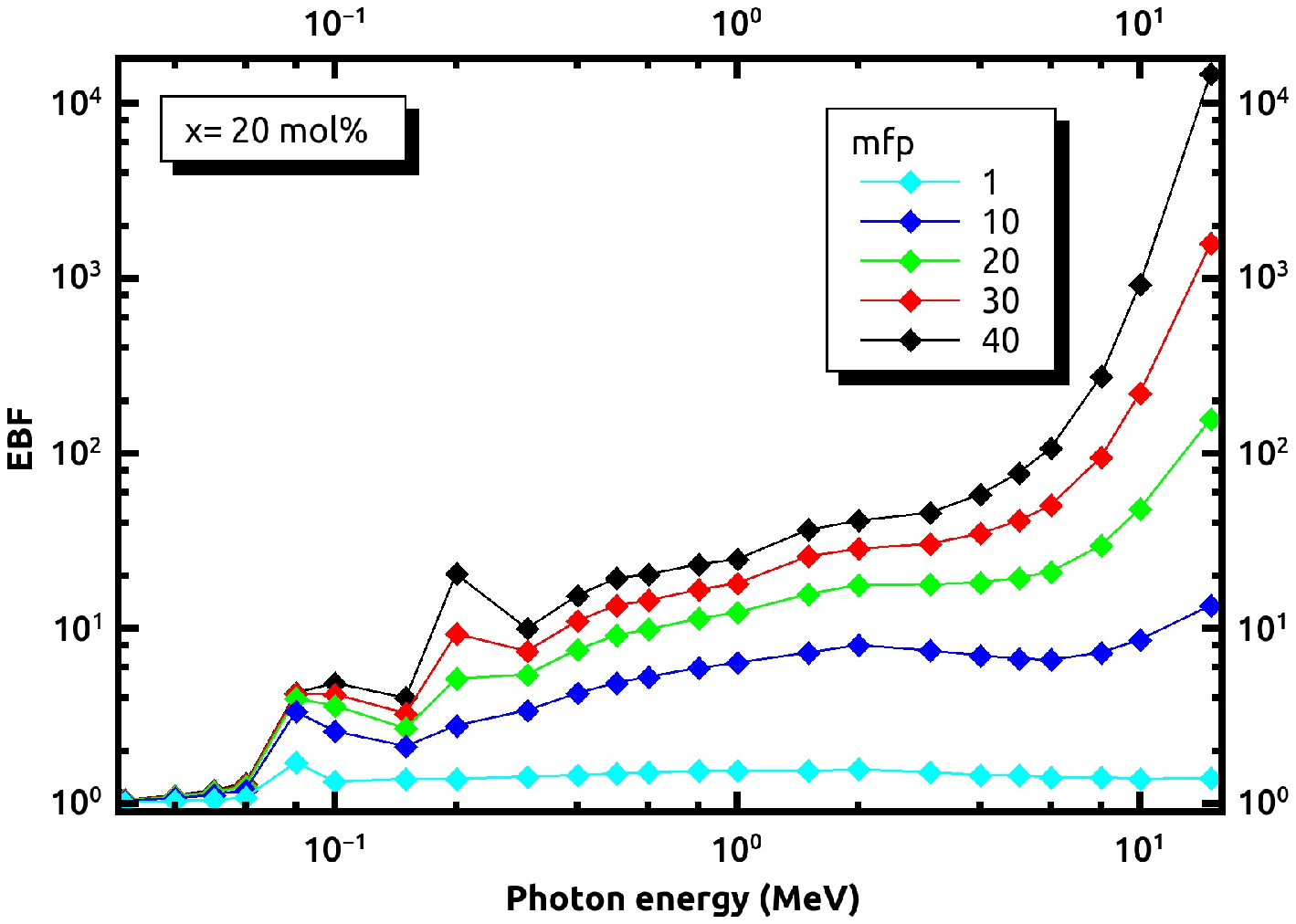}
  
  \label{fig:sub7}
\end{subfigure}%
\begin{subfigure}{.6\textwidth}
  \centering
  \includegraphics[width=1.08\linewidth]{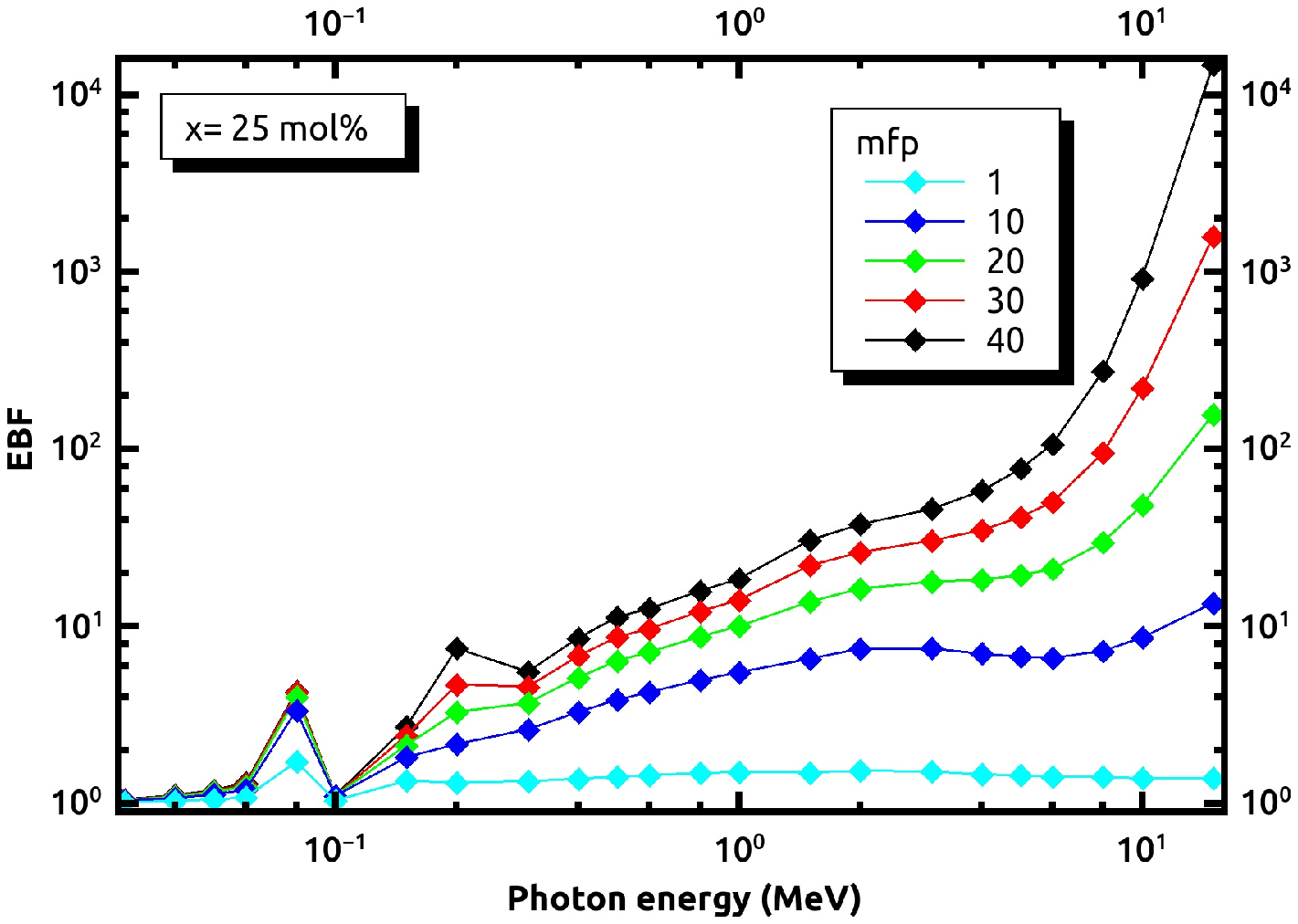}
  
  \label{fig:sub8}
\end{subfigure}
\caption{Exposure buildup factors as a function of photon energy for the borate-based  glasses with different molar fractions of $Bi_2O_3$.}
\label{fig6}
\end{figure}

\begin{figure}[H]
\centering
\begin{subfigure}{.6\textwidth}
  \centering
  \includegraphics[width=1.08\linewidth]{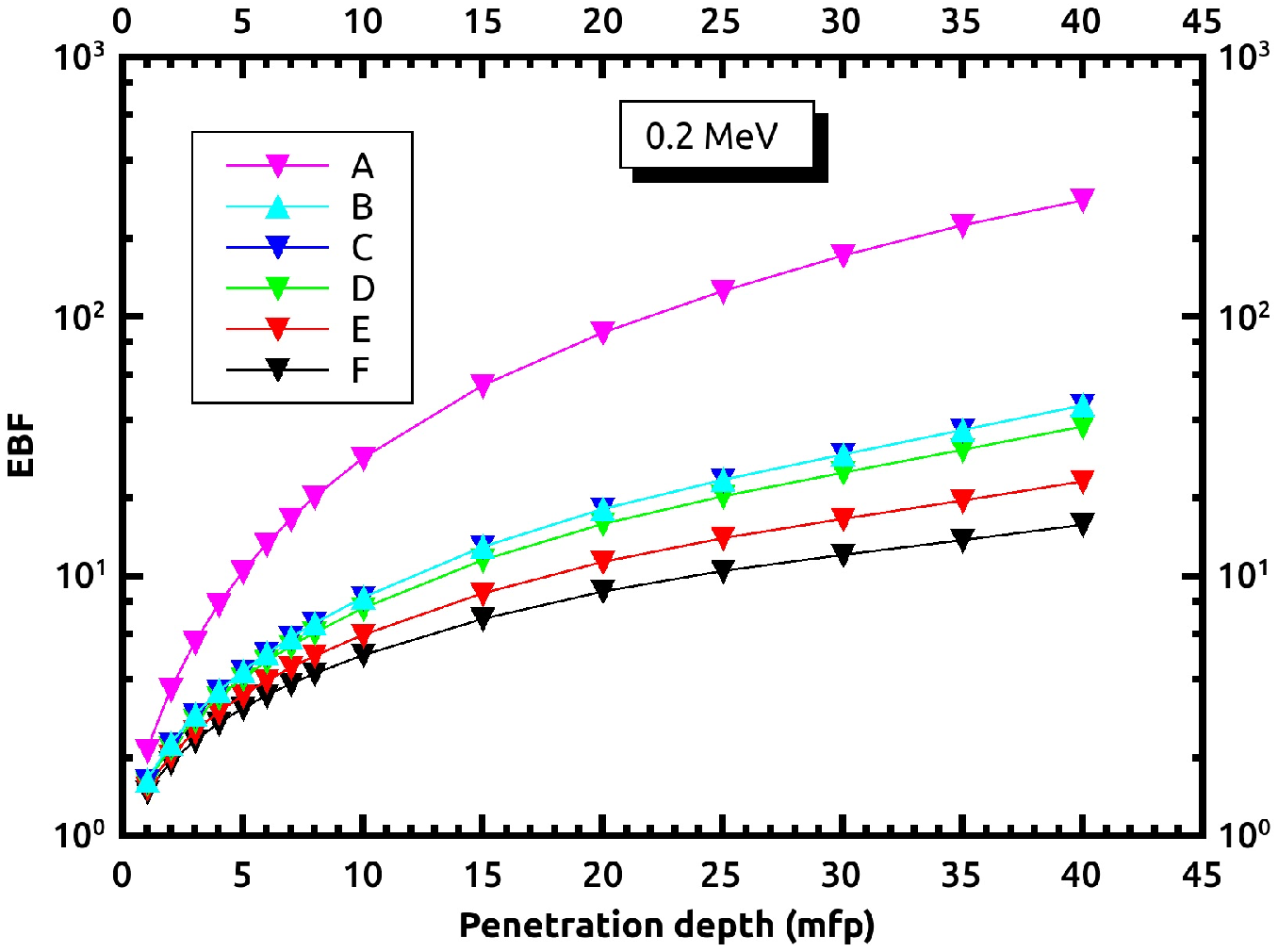}

  \label{fig:sub9}
\end{subfigure}%
\begin{subfigure}{.6\textwidth}
  \centering
  \includegraphics[width=1.08\linewidth]{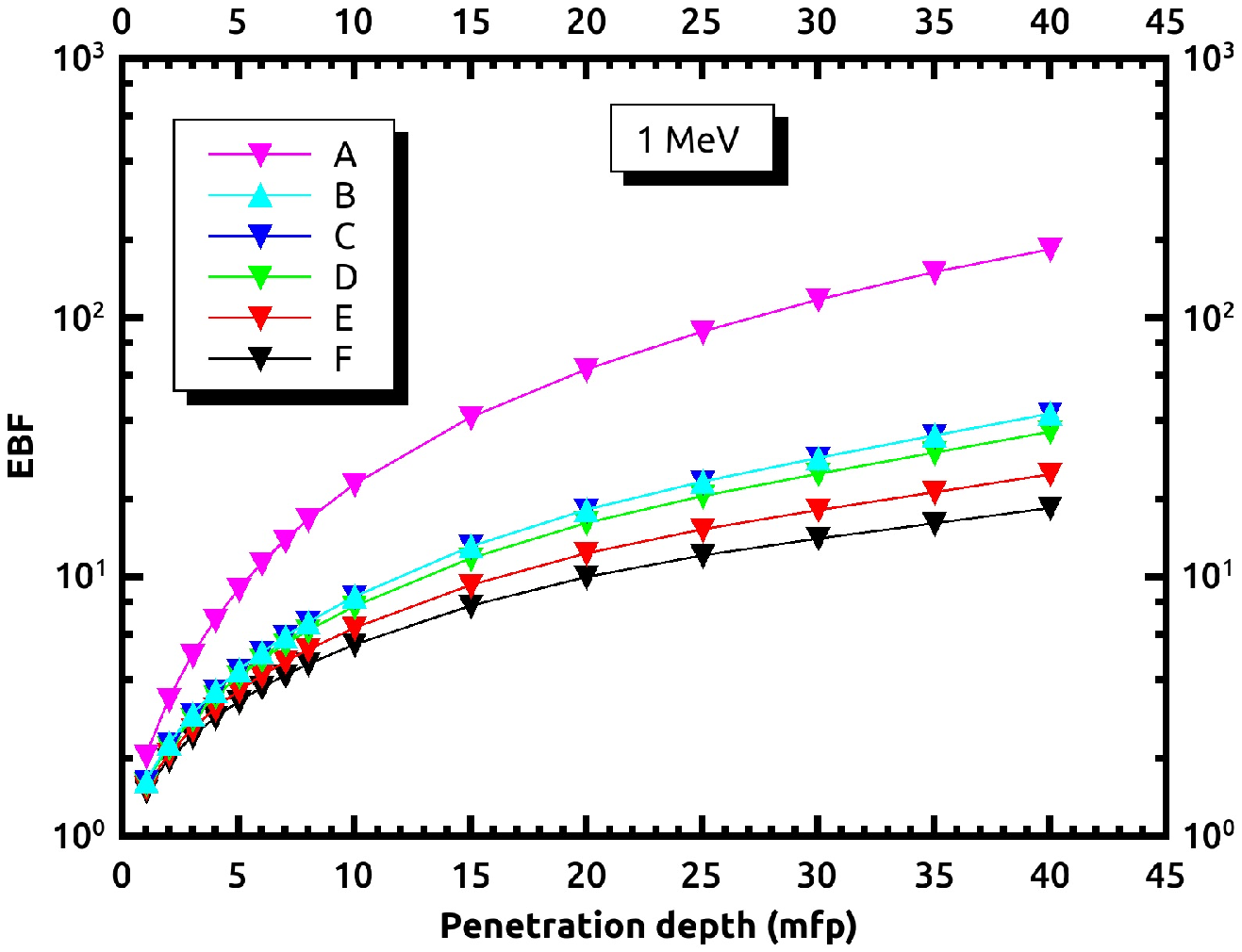}
  
  \label{fig:sub10}
\end{subfigure}
\begin{subfigure}{.6\textwidth}
  \centering
  \includegraphics[width=1.08\linewidth]{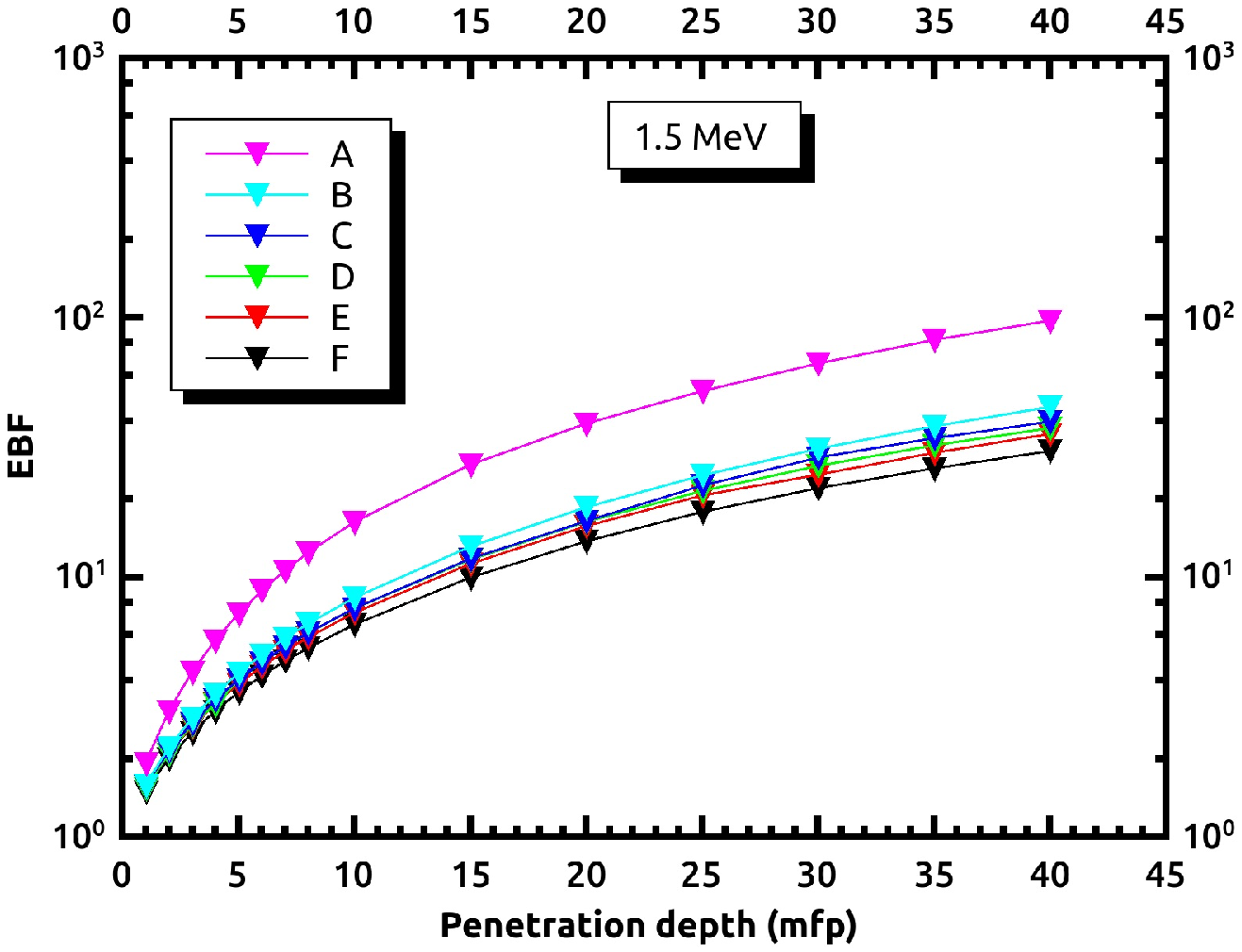}
  
  \label{fig:sub11}
\end{subfigure}
\caption{Exposure buildup factors of the borate-based  glasses with penetration depth (mfp) at photon energies 0.2, 1 and 1.5 MeV.}
\label{fig7}
\end{figure}

\begin{figure}[H]
\begin{center}
\includegraphics[width=0.8\textwidth,natwidth=610,natheight=642]{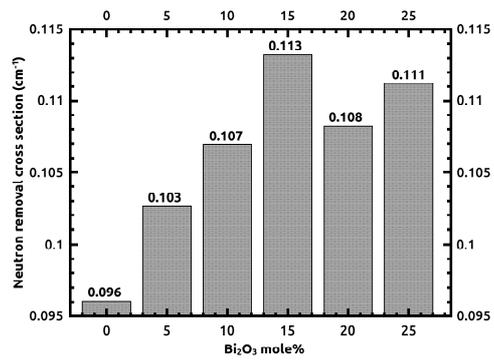}
\caption{Effective removal cross section of fast neutrons for borate-based  glasses.}
\label{fig8}
\end{center}
\end{figure}

Fig.~\ref{fig3} shows the results of $Z_{eff}$ with photon energy (between 0.001 MeV to 10 MeV) for the studied glass samples. It is clearly seen that the values of $Z_{eff}$ increase from the glass A to the glass F over the considered energy range. Such that the order of $Z_{eff}$ values is : A$<$B$<$C$<$D$<$E$<$F. This is attributed to increase the concentration of $Bi_2O_3$ from glass A (with 0 mole$\%$) to glass F (with 25 mole$\%$). In the fact, the good shielding materials should have a higher $Z_{eff}$. For example, the gamma ray interact more with high $Z_{eff}$ materials leading to decrease the photon energy that becomes incapable to penetrate the material. Thus, the glass F is a better shielding material among our glass samples. Also, as shown in Fig.~\ref{fig3}, the $Z_{eff}$ varies with photon energy for a given glass sample. Such that the $Z_{eff}$ values increase at low photon energies, decrease at intermediate photon energies and again increase at high photon energies. The reason is that the partial photon processes are related to the Z of the material. The photoelctric process depends on $Z^4$, the Compton process depends on Z, and the pair production process depends on $Z^2$. Moreover, when the photon energy equals the electron binding energy of K, L, or M shells, an atypical change of the attenuation properties (e.g. $\mu_m$ or $Z_{eff}$) was observed around those energies. This phenomenon can be seen clearly in the present glass samples due to the electron binding energies in Bi is relatively high, for example it is 90.5 keV for the K-shell. The half value layer (HVL) for borate-based glasses in the energy range from 0.001 MeV to 10 MeV, is shown in Fig.~\ref{fig4}. It is clearly that the HVL values increase sharply with increasing the photon energy up to 3 MeV. Thereafter, the values of HVL show a slow increase with the increasing the photon energy up to 10 MeV. From Fig.~\ref{fig4}, it is also observed that the HVL values decrease as the increase of $Bi_2O_3$ content. So, we found that the glass F (with the highest $Bi_2O_3$ content) has the lowest values of HVL, and this refers to the efficiency of the glass F to use as the shielding material. The mean free path (MFP) for borate-based  glasses with respect to the concentration of $Bi_2O_3$ at different photon energies, is shown in Fig.~\ref{fig5} as; (a) a comparison with ordinary concretes and (b) a comparison with other glasses. From these figures, the values of MFP were found to be lower as $Bi_2O_3$ content increases indicating that the shielding features are improved. From Fig.~\ref{fig5}(a), It were found that the shielding efficiency properties of our studied glass samples are better than those of ordinary concretes, for example the values of MFP for our studied glass samples are lower than those for the ordinary concretes. Also, in Fig.~\ref{fig5} (b), we compared our glass samples with other reference glass samples: the Glass 1 (75HPSg -- 20Na2O -- 5$Bi_2O_3$) at photon energies of 1.332, 1.173 and 0.662 MeV, the Glass 2 (55BiO3 -- 54B2O3) at photon energies of 1.332 and 0.662 MeV, and the Glass 3 (15B2O3 -- 50$Bi_2O_3$ -- 15ZnO -- 20Li2O) at photon energy of 0.300 MeV. It is clear that the present glass samples with $Bi_2O_3$ $\geq$ 10 are found to be better than the Glass 1 in terms of shielding efficiency. However, they have lower shieding efficiency than the Glass 2. Moreover, the present glass samples with $Bi_2O_3$ $\geq$ 20 mol$\%$ show an equivalent shielding efficiency for the Glass 3. It should to be mentioned that the Glass 2 (55 mol$\%$ of $Bi_2O_3$) and the Glass 3 (50 mol$\%$) have a concentration of $Bi_2O_3$ more than the studied glasses (25 mol$\%$). This makes the present glasses more economical than the Glass 2 and the Glass 3 (e.g. low amount of $Bi_2O_3$). Fig.~\ref{fig6} shows the values of EBF in the energy region 0.015 -- 15 MeV for all the studied glasses at different penetration depth 1, 10, 20, 30 and 40 mfp. (EBF, calculated as  explained previously ~\cite{al2019radiation}). In the case of x= 0 mol$\%$ (the absence of Bi from the glass sample), it can be noted that the values of EBF increase gradually in the low energy region with increasing the photon energy and the penetration depth. After that there is a peak in the intermediate energy region, then the EBF values decrease in the high energy region with increasing the photon energy and still increase regularly as the penetration depth increases. This is attributed to the dependence of the partial photon processes and it is similar to the findings that were reported for different glass systems modified by high-Z element (e.g. Bi in the present work) such as Pb. In the cases of 5 $\leq$ x $\leq$ 25 mol$\%$, the photoelectric effect dominates at low energies and does not allow the photons to buildup in the medium. Therefore, the small values of EBF have been observed in this region. With increasing the photon energy, the EBF values increase due to the multiple scattering (e.g. Compton scattering). At high energies around 10 MeV, the pair production is the dominating process and the swift increase in the values of EBF may be attributed to that the annihilation of the electron--positron pair originates secondary photons which likly buildup for large penetration depths. The variation of EBF with penetration depth at different energies 0.2, 1 and 1.5 MeV is shown in Fig.~\ref{fig7}. It is clear that the EBF values increase as penetration depth increases due to the increment of the photons number in the medium from the scattering and the pair annihilation processes. Also, from Fig.~\ref{fig7}, it is obvious that the concentration of $Bi_2O_3$ plays an important role to decrease the EBF values at low energies, for example the glass F has the lowest values of EBF at 0.2 MeV. This phenomenon becomes trivial with increasing the photon energy (see Fig.~\ref{fig7} at 1.5 MeV). The reason is that at energies $>$ 1.022 MeV the absorption processes are too scarce. The removal cross section of fast neutrons ($\sum {_R}$, calculated as  explained previously ~\cite{al2019radiation}) for the studied glasses is shown in Fig.~\ref{fig8}. The obtained results of $\sum {_R}$ are varied between 0.096 $cm^{-1}$ and 0.113 $cm^{-1}$. The glass D (with 15 mol$\%$ $Bi_2O_3$) has the highest value of $\sum {_R}$ among the studied glass samples, so it is a good candidate for fast neutron shielding.

\section{Conclusion}

In summary, the nuclear shielding properties of the borate-based glass systems in the form of $(75-x)B_2O_3-xBi_2O_3-10Na_2O-10CaO-5Al_2O_3$ (0 $\leq$ x $\leq$ 25 mol$\%$) have been reported to serve for gamma-ray and neutron protection applications. The $\mu_m$ values of the selected glasses were computed by using both XCOM and Geant4 simulation codes. It is found that the  $\mu_m$ values of the borate-based glass vary with changing the photon energy and the amount of $Bi_2O_3$ content. The results generated by XCOM and Geant4 simulation codes are almost overlapping. For shielding from gamma and neutrons radiation, the selection of an appropriate glass can be done based on data of Z$_{eff}$, HVL and MFP, consideration of neutron removal cross sections, exposure build-up factor and also, according to desired application. It can be concluded that the addition of $Bi_2O_3$ has played a crucial role in protection ability for the glass F against gamma rays, while the glass D offers the best shielding features against the fast neutrons. This work on shielding properties of the borate-based glasses would be useful in various radiation protection applications such as exposure control.

%\begin{acknowledgements}
%\end{acknowledgements}

\bibliographystyle{spphys}       % APS-like style for physics
\bibliography{mybibfile}   % name your BibTeX data base

 \begin{table}[H]
 \caption{Density and chemical composition of the selected glasses.}
{\begin{tabular}{l@{\hskip 0.5in}l@{\hskip 0.5in}llllll}
\toprule \toprule
Sample     & Density $(g/cm^3)$   & \multicolumn{5}{c}{ Glass compositions (mol$\%$)}  \\
\cmidrule(lllll){3-7}
 &  &$B_2O_3$ &$Bi_2O_3$&$Na_2O$&$CaO$&$Al_2O_3$ \\
\toprule
A       & 2.250  &75 & 0 & 10& 15&  5  \\
 
B     &  2.938     & 70 & 5 & 10& 15& 5 \\

C     & 3.572    &65 & 10 & 10& 15& 5\\

D      &  4.269  & 60 & 15 & 10& 15& 5 \\

E     &   4.509 & 55 & 20 & 10& 15& 5 \\

F    &  5.035   & 50 & 25 & 10& 15& 5\\

\bottomrule \bottomrule
\end{tabular}}
\label{tab1}
\end{table}

\end{document}